\documentclass{PoS}

\usepackage[pdftex]{graphicx}
\usepackage[dvips]{epsfig}

\usepackage{bbm}
\usepackage{verbatim}
\usepackage{psfrag}
\usepackage{slashed}
\usepackage[normalem]{ulem}

\usepackage{amsmath}
\usepackage{amssymb}
\usepackage{marvosym}
\usepackage{stmaryrd}
\usepackage{wasysym}
\usepackage{mathcomp}
\usepackage{multirow}

\def \Nc {{N_{c}}}
\def \Nf {{N_{f}}}

\newcommand{\expval}[1]{\langle #1 \rangle}

\newcommand{\lr}[1]{\left( #1 \right)}

\newcommand{\beqn} {\begin{equation}}
\newcommand{\eqn} {\end{equation}}

\def \beq{\begin{equation}}
\def \eeq{\end{equation}}
\def \bea{\begin{eqnarray}}
\def \eea{\end{eqnarray}}

\def \Tr {{\rm Tr}}

\def \bet0{\beta_0}
\def \bet1{\beta_1}
\def \simgt{\,\rlap{\lower 7.5 pt\hbox{$\mathchar \sim$}}\raise 3 pt \hbox{$>$}\,}
\def \simlt{\,\rlap{\lower 7.5 pt\hbox{$\mathchar \sim$}}\raise 3 pt \hbox{$<$}\,}

\def\lsim{\raise0.3ex\hbox{$<$\kern-0.75em\raise-1.1ex\hbox{$\sim$}}}
\def\gsim{\raise0.3ex\hbox{$>$\kern-0.75em\raise-1.1ex\hbox{$\sim$}}}

\def \pbp {\bar{\psi} \psi}

\title{A surprise with many-flavor staggered fermions\\ in the strong coupling limit}

\ShortTitle{Many-flavor staggered fermions in the SC-limit}

\author{Philippe de Forcrand\\
Institut f\"ur Theoretische Physik, ETH Z\"urich, CH-8093, Switzerland\\
CERN, Physics Department, TH Unit, CH-1211 Geneva 23, Switzerland  \\
        E-mail: \email{forcrand@phys.ethz.ch}}

\author{Seyong Kim\\
        Department of Physics, Sejong University, Seoul 143-747, Korea \\
        E-mail: \email{skim@sejong.ac.kr}}
\author{\speaker{Wolfgang Unger}\\
      Institut f\"ur Theoretische Physik, Goethe-Universit\"at Frankfurt, 60438 Frankfurt am Main, Germany\\
       E-mail: \email{unger@th.physik.uni-frankfurt.de}}

\abstract{
\vspace*{-12.5cm}
\begin{flushright}
\texttt{\footnotesize CERN-PH-TH/2012-302}\hspace{-7mm}\vphantom{.}\\
\end{flushright}
\vspace*{11.5cm}
It is widely believed that chiral symmetry is spontaneously broken at zero temperature in the strong coupling limit of staggered fermions, for any number of colors and flavors. 
Using Monte Carlo simulations, we show that this conventional wisdom, based on a mean-field analysis, is wrong. 
For sufficiently many fundamental flavors, chiral symmetry is restored via a bulk, first-order transition. 
This chirally symmetric phase appears to be analytically connected with the expected conformal window of many-flavor continuum QCD. 
We perform simulations in the chirally symmetric phase at zero quark mass 
for various system sizes $L$, 
and measure the torelon mass and the Dirac spectrum. 
We find that all observables scale with $L$, which is hence the only infrared length scale.
Thus, the strong-coupling chirally restored phase appears as a convenient 
laboratory to study IR-conformality. Finally, we present a conjecture for the
phase diagram of lattice QCD as a function of the bare coupling and the number
of quark flavors.}

\FullConference{The 30 International Symposium on Lattice Field Theory - Lattice 2012,\\
		June 24-29, 2012\\
		Cairns, Australia}

\begin{document}


\section{Introduction}
The possibility that the Higgs boson could be a composite bound-state in
a high-energy Technicolor theory, together with the requirement that such a theory
be ``walking'' in order to accommodate stringent bounds on 
flavor-changing neutral currents, has motivated
several large-scale computer simulations. 
To determine via Monte Carlo simulations whether a given theory with specific gauge group and fermion content is 
inside the conformal window, i.e. possesses an infrared fixed point (IRFP),
is particularly challenging:
one must probe 
the extreme infrared properties of the theory, while at the same time
taking the continuum limit, and controlling
the chiral limit. Due to these difficulties, in spite of considerable efforts, there is no consensus yet
on the minimum number $\Nf^*$ of fundamental quark flavors needed for QCD to be inside the
conformal window~\cite{review}.\footnote{A numerical demonstration of walking has been provided
 in the 2-$d$ $O(3)$ model at vacuum angle $\theta\approx\pi$~\cite{O3}.}

Here, we relax the demand that results should be obtained in the continuum
limit. On a coarse lattice, long-distance properties can be studied more
economically. While such properties may differ from those of the 
corresponding continuum theory, it is still instructive to consider
the possible existence of an IRFP for a discretized lattice theory.
The phase diagram of SU$(N_c)$ gauge theory with $N_f$ fundamental fermions,
as a function of $N_f$ and the bare gauge coupling,
has been predicted in the celebrated Ref.~\cite{Miransky-Yamawaki}. It is important to confront these
predictions with uncontroversial numerical evidence.
Therefore, we start our investigation by considering the strong coupling
limit, where the lattice is maximally coarse.\footnote{ 
Note that we consider standard staggered fermions, and the standard plaquette action. 
Other discretizations could lead to different results, 
only the continuum limit is universal.}
The conventional wisdom for strong coupling QCD with staggered fermions is that chiral symmetry always remains broken at zero temperature, 
regardless of the values of $\Nf$ and $\Nc$. This belief is based on 
mean-field analyses performed in some of the earliest papers on lattice QCD.
In particular, it was shown in \cite{Kluberg1983} that at leading order in 
a $1/d$ expansion, the chiral condensate has a value independent of $\Nf$ and $\Nc$, 
and depends only on the number $d$ of spatial dimensions, $\expval{\pbp}=\sqrt{\frac{2}{d}}\left(1-\frac{1}{4d}\right)$.
Chiral symmetry may be restored by increasing the temperature $T$,\footnote{We find, following the approach of \cite{Damgaard1985}, 
that $T_c$ is non-zero for all $\Nf$: $aT_c=\frac{d}{4}+\frac{d}{8}\frac{\Nc}{\Nf}+\mathcal{O}\lr{\frac{1}{\Nf^2}}$}
but will never be restored at $T=0$.
Since naively mean-field theory is expected to work well when the number of d.o.f. per site 
is large, there was no reason to doubt this result.
Besides, it is in accord with the intuition that the gauge field is maximally
disordered in the strong-coupling limit, and the disorder drives
chiral symmetry breaking.
On the other hand, one might expect the above disorder to be modified 
by dynamical fermions, which have an ordering effect. 
Indeed, the loop expansion of the determinant shows that the fermionic 
effective action 
$S_{\rm eff} = -\log\det(\slashed{D} + m_q)$ starts with a positive 
plaquette coupling $\Delta\beta$, proportional to $1/m_q^4$ for heavy quarks \cite{Hasenfratz1994}. 
Clearly, for $\Nf$ flavors, $S_{\rm eff} \propto \Nf$, and hence $\Delta\beta \propto \Nf$.
This plaquette term suppresses fluctuations in the gauge field, 
which suggests that chiral symmetry restoration might take place for sufficiently large $\Nf$.

\section{Monte Carlo results}

\begin{figure}[t!]
\centerline{
\includegraphics[width=0.45\textwidth]{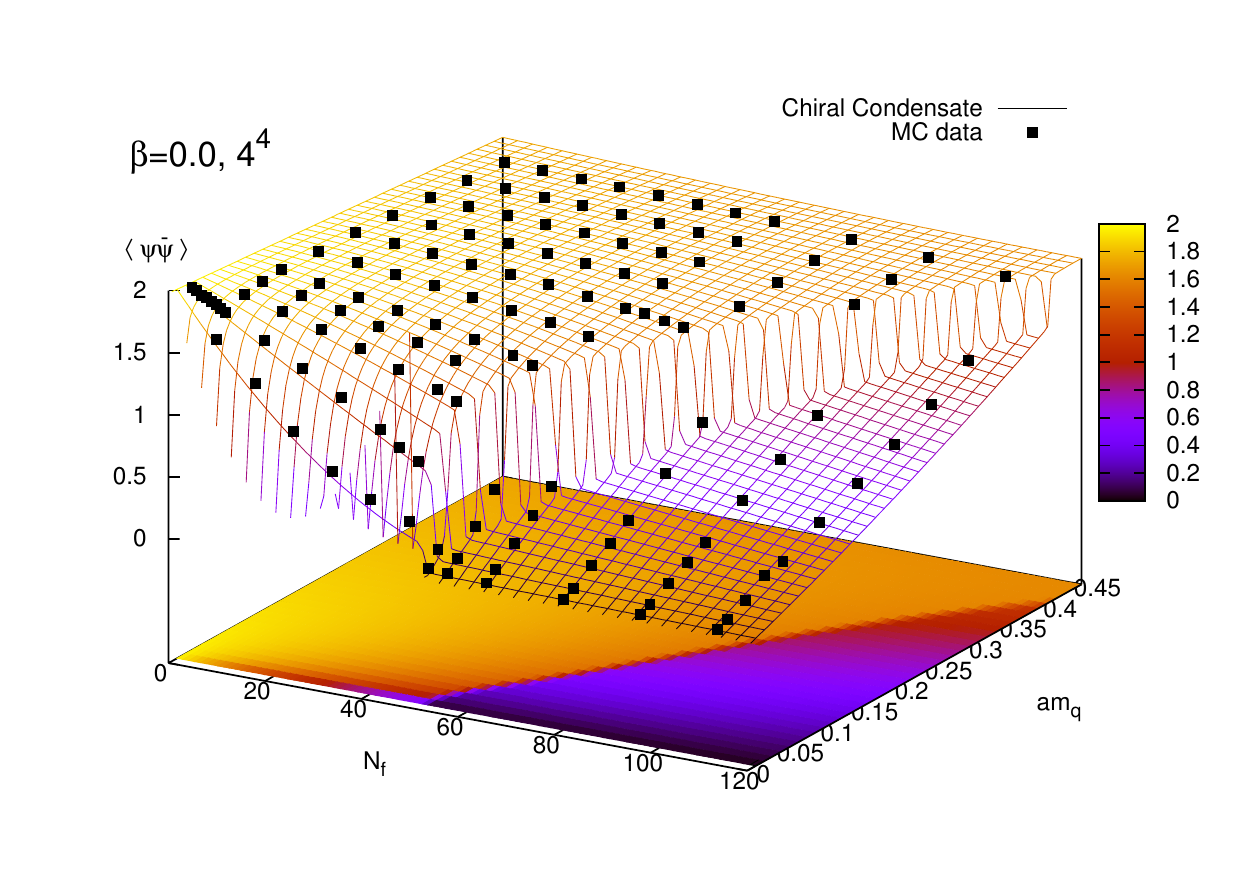}
\includegraphics[width=0.45\textwidth]{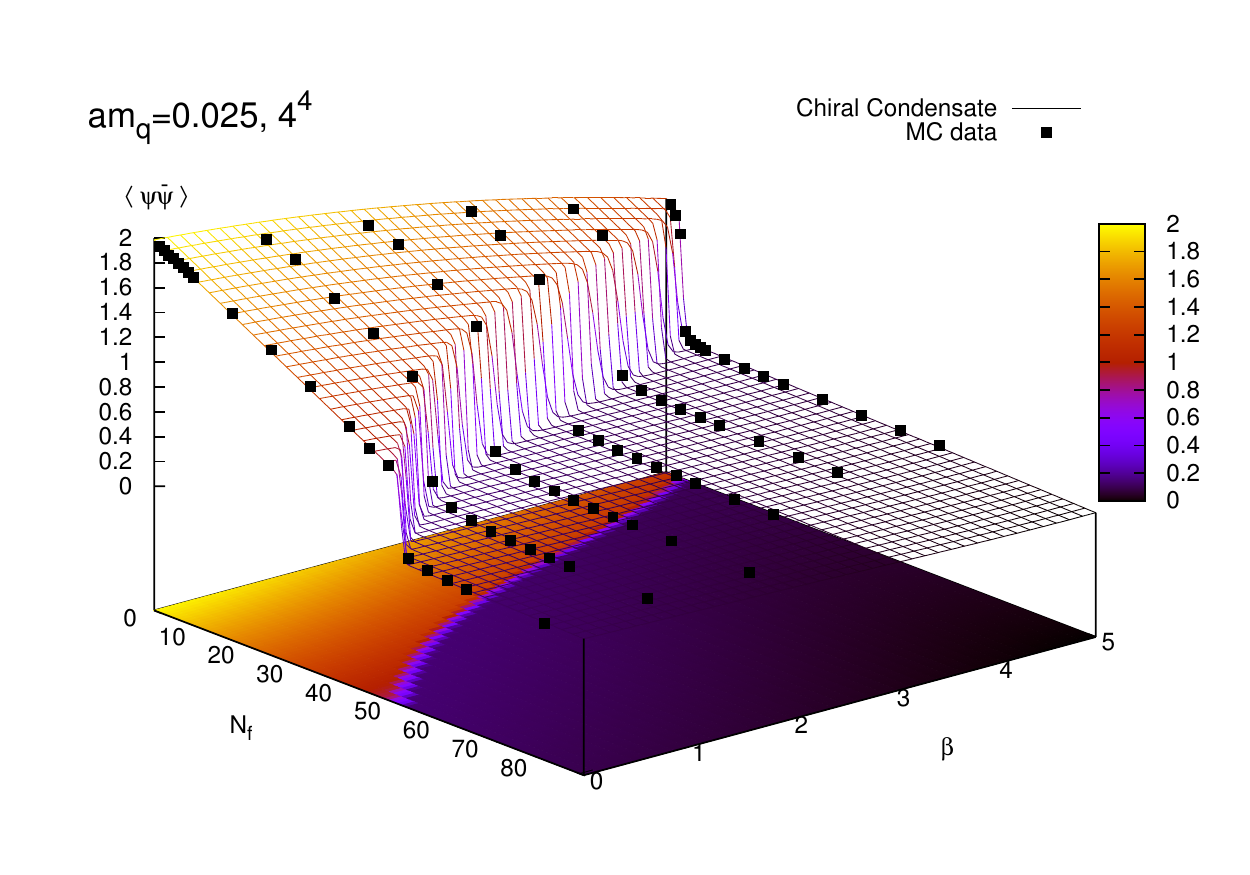}
}
\caption{The chiral condensate, left: at strong coupling, $\beta=0$, in the $(\Nf,am_q)$ plane, for $4^4$. Right: at $am_q=0.025$ in the $(\Nf,\beta)$ plane, for 
$4^4$. We have compared with larger volumes \cite{JHEP} and find that the phase transition remains strongly 
first order at weaker coupling.
The contour plot indicates the qualitative behaviour of the phase boundary, extending smoothly to smaller $\Nf$ at weaker coupling.}
\label{fixedam}
\label{beta0}
\end{figure}

The only way to resolve this puzzle is to perform Monte Carlo simulations 
in the strong coupling limit of staggered fermions, to detect a possible chiral 
symmetry restoration for sufficiently large $\Nf$. 
These simulations are straightforward, using the standard HMC
algorithm.
As expected, the effect of increasing $\Nf$ on the chiral condensate is to 
reduce its magnitude. But it came as a surprise to find that the chiral condensate vanishes 
via a strong first-order transition at $\Nf^c\simeq 52$ continuum flavors in the chiral limit (i.e. $\Nf^c/4 \simeq 13$ staggered fermion fields).
In the broken phase, $\expval{\pbp}$ remains almost constant. It vanishes 
in the chiral limit due to finite-size effects only. 
In contrast, in the chirally restored phase the condensate is caused by
explicit symmetry breaking and is proportional to the quark mass.
This is illustrated Fig.~\ref{beta0} (left), where the condensate is shown
as a function of $\Nf$ and bare quark mass $(a m_q)$.
This $\Nf$-driven transition turns out to be a bulk, zero-temperature
transition: finite-size effects on the phase boundary are small when 
comparing two different system sizes $4^4$ and $6^4$.
The critical number of flavors increases with the quark mass.\footnote{
Heavier quarks have a weaker
ordering effect, so that the induced plaquette coupling $\Delta\beta$ 
decreases if one keeps $\Nf$ fixed. It takes more flavors to keep the
system chirally symmetric: $\Nf^c\propto (a m_q)^4$.}

Having established an $\Nf$-driven phase transition in the strong-coupling
limit, we may consider its impact on the lattice theory at non-zero
lattice gauge coupling $\beta$ as well.
Since the transition is strongly first order, it has to persist for some range
in $\beta$ at least.
In fact, we find a smooth variation of the $\Nf$-driven transition with $\beta$ at a given small quark mass $am_q=0.025$, as shown Fig.~\ref{fixedam} (right). 
The transition extends to weak coupling, at least to $\beta=5$, and remains 
strongly first-order. Thus, it is plausible that this transition, which
separates a chirally broken (small $\Nf$) and a chirally symmetric (large $\Nf$)
phase, persists all the way to the $\beta\to\infty$ continuum limit, where
it is to be identified with the transition at $\Nf=\Nf^*$ between the
chirally broken and the IR-conformal, chirally symmetric phase. 
In other words, our chirally restored phase may be analytically connected 
to the conformal window in the continuum limit, because
we do not observe any additional non-analyticity as $\beta$ is increased.
This possibility motivates our study of the properties of the strong-coupling
chirally symmetric phase, looking for tests of IR-conformality.

\section{Looking for conformality in the chirally symmetric phase}

\newcommand{\rank}{\text{rank}}

It is natural to ask whether the chirally restored phase is connected to the conformal window, i.e.~whether the chirally restored phase at strong coupling is IR-conformal, and if so, 
whether it is trivial (IR fixed point coupling $g^*=0$).
In this section we present measurements of gluonic and fermionic observables
chosen to address these
questions. Our results support the following conclusion: the strong-coupling chirally 
symmetric phase is IR-conformal, and it is non-trivial.
The simulations performed here are all in the chirally symmetric phase 
at zero plaquette coupling, with $N_f=56$ and $96$ continuum flavors, 
and with lattices of size $4^3\times 16$ to $12^3\times 24$.
The quark mass is set exactly to zero, which is computationally feasible because the Dirac operator has a spectral gap in the symmetric phase.
Moreover, having one infrared scale, the system size $L$,
rather than two scales ($L$ and $1/m_q$) is of great advantage when analyzing
the results.\footnote{
Note that the average plaquette values we measure, 
$\approx 0.35$ $(N_f=56)$ and $\approx 0.52$ $(N_f=96)$, are very far from 0 ($\Nf=0$, maximally disordered): the ordering effect of the dynamical fermions plays a dominant
role, and the vanishing of the plaquette coupling is not associated with special properties.}

\begin{figure}
\centerline{
\includegraphics[width=0.45\textwidth]{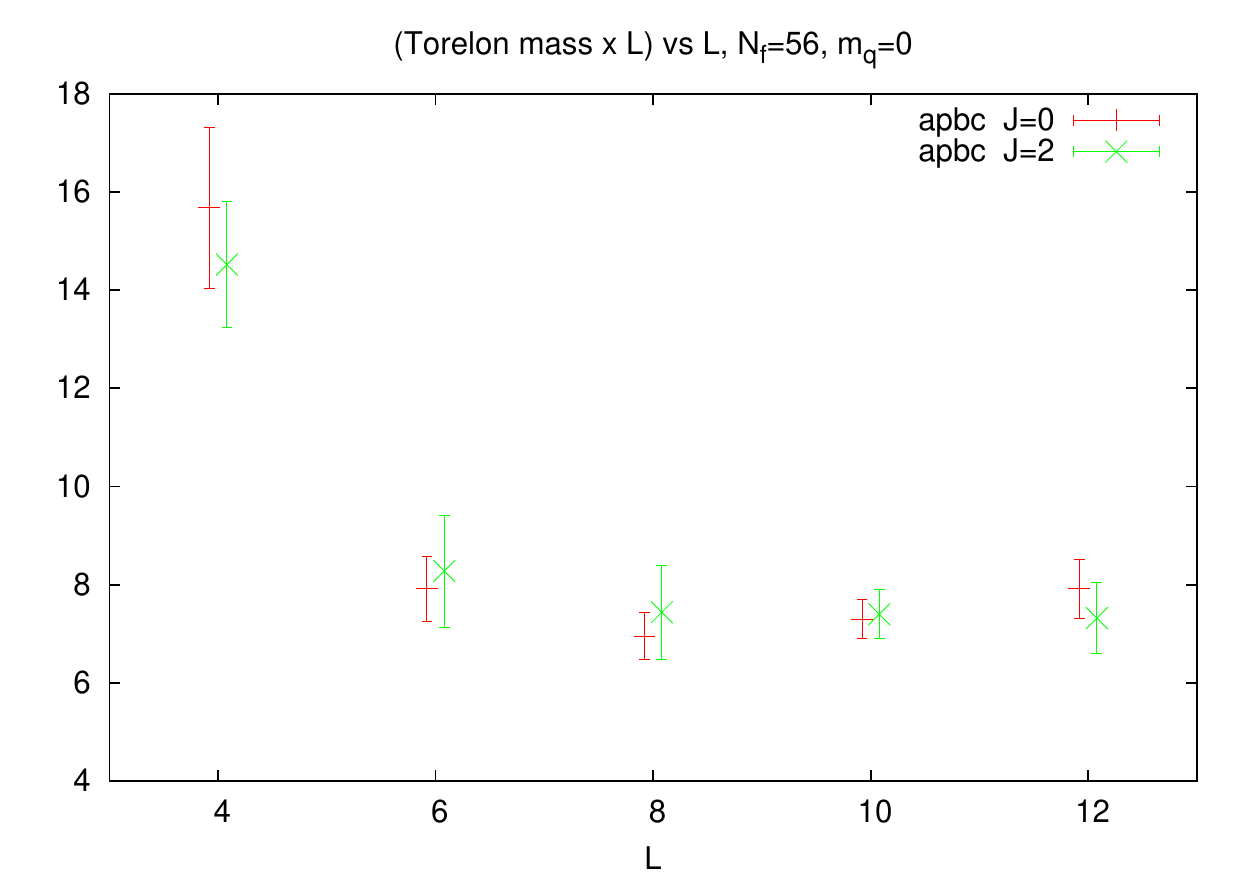}
\includegraphics[width=0.45\textwidth]{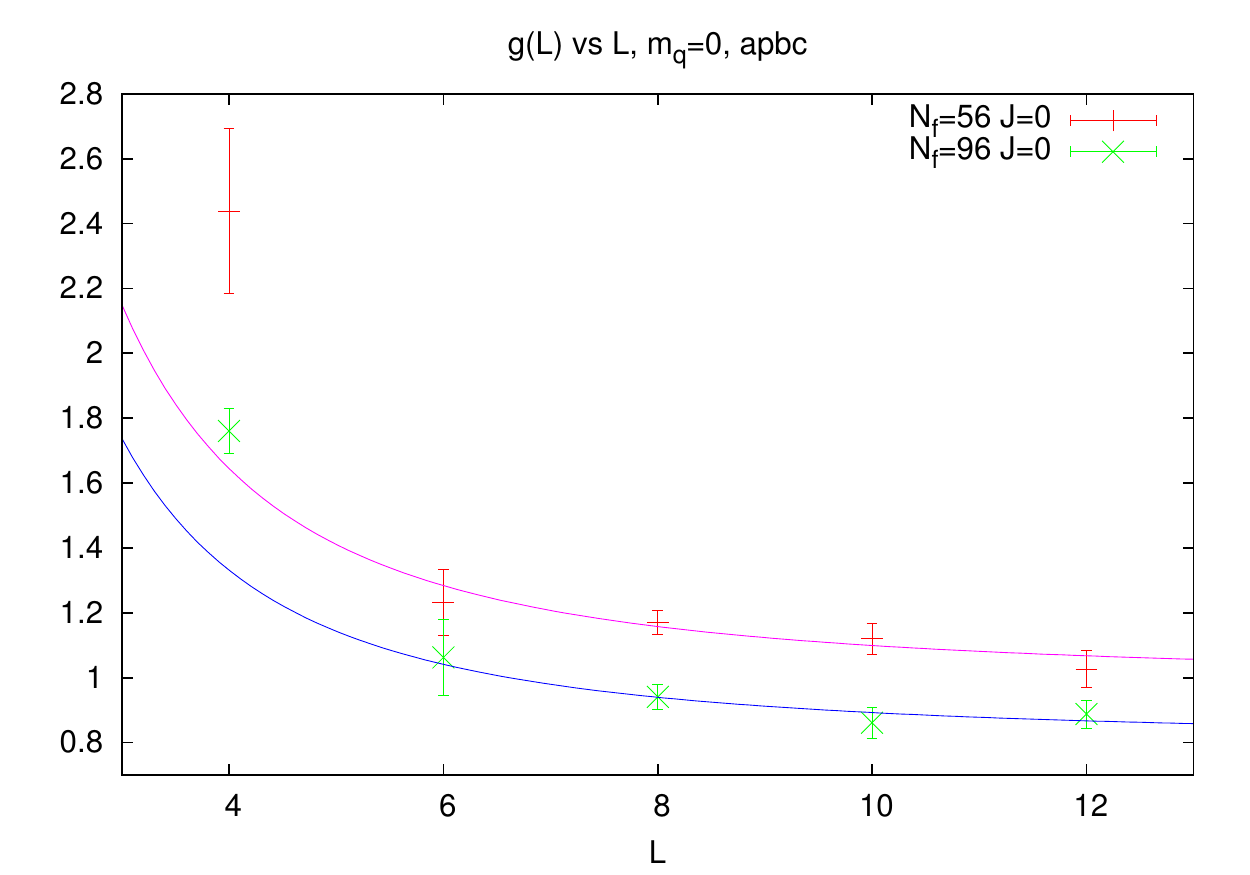}
}
\caption{Left: The torelon mass $m_T(L)$ multiplied by $L$ versus $L$ for $\Nf=56$ (for anti-periodic b.~c.), which is constant for large enough $L$.
Right: The running coupling $g(L)$ defined via the temperature-dependence of 
the Debye mass, which is identified with the torelon mass, for $N_f=56$ and 96.
For each $N_f$, the curve is a fit to a constant plus $(a/L)^2$ corrections of 
the 4 largest volumes. The larger $N_f$ has a smaller coupling.}
\label{torelon}
\label{running_g}
\centerline{
\includegraphics[width=0.45\textwidth]{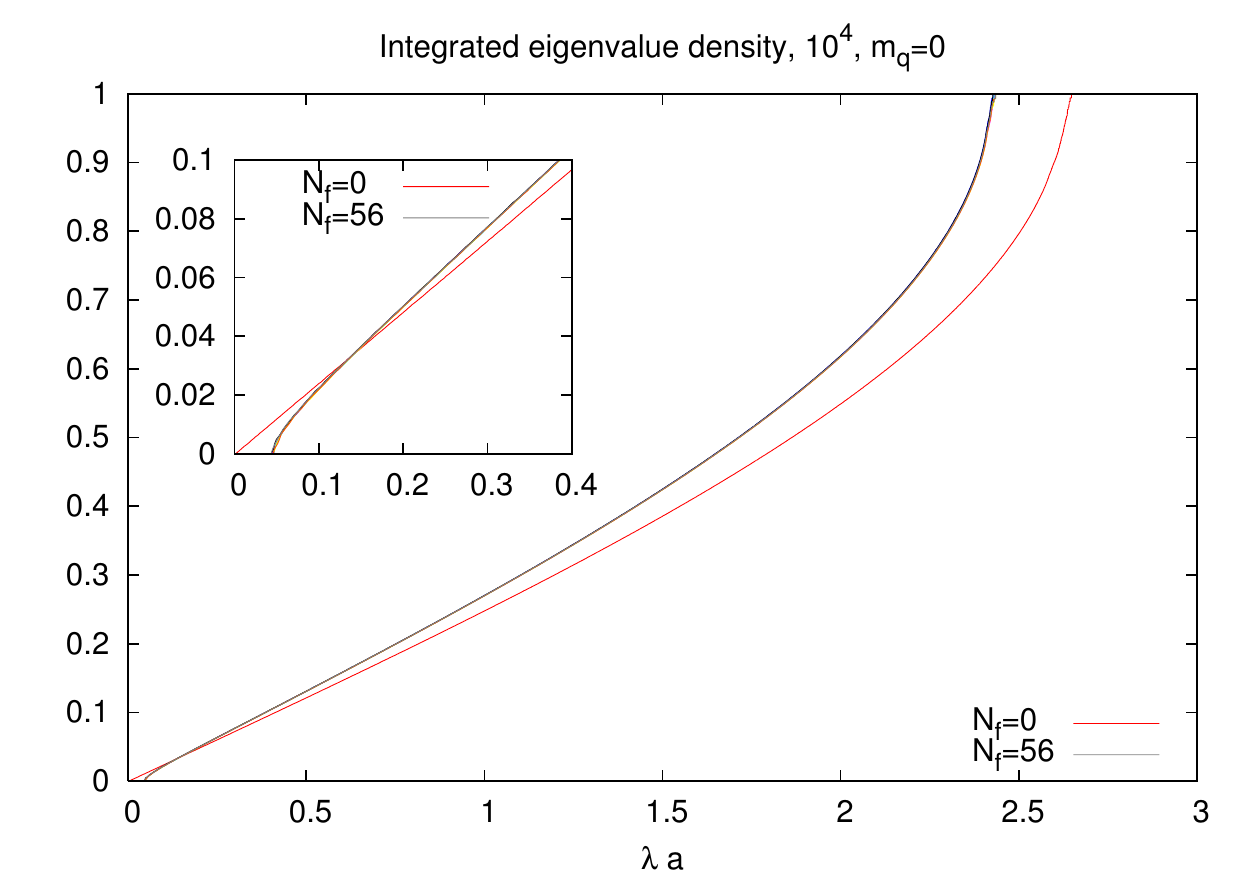}
\includegraphics[width=0.45\textwidth]{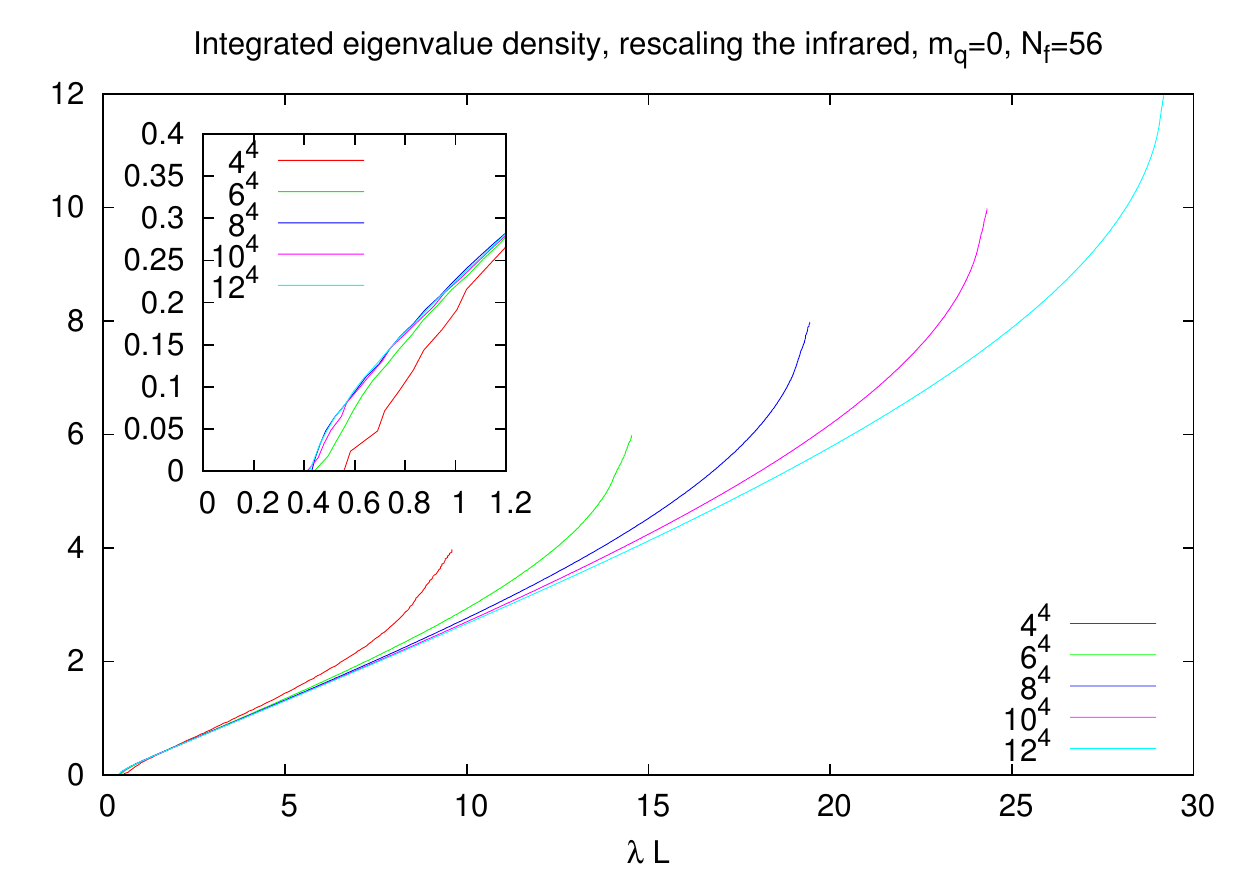}
}
\caption{The integrated eigenvalue density. Left: comparison of $\Nf=0$ (quenched) with $\Nf=56$ in the chirally restored phase, where only the latter shows a spectral gap.
Right: The rescaled spectral gap, indicating $1/L$ scaling.
10 configurations are superimposed, variations in the spectrum are very small.
}
\label{dirac}
\end{figure}

$\bullet$ The ``torelon'' is a gluonic excitation which is topologically non-trivial:
it is excited by any \linebreak Wilson line which wraps around the spatial boundary
in one spatial direction $i=1,2,3$:\linebreak
$T_i(t) = \Tr \prod_{k=0}^{L-1} U_i(\vec{x}+k\hat{i},t)$.
We extract the mass of this excitation from 
the exponential decay of the correlator $\langle T_i(0)^* T_i(t) \rangle$.
To suppress excited states, we smear the links within each time-slice
before constructing $T_i$. This observable has been used for a long time
to extract the string tension $\sigma$ in Yang-Mills theories~\cite{Michael1989}:
it can be viewed as a loop of gluonic string, whose energy $m_T(L)$ grows with
its length as $\sigma L$. Our naive expectation that $m_T(L)$ grows with $L$ until string breaking occurs
was however disproved by our measurements: the dimensionless quantity $a m_T(L)$ {\em decreases} on larger lattices corresponding
to a larger ratio $L/a$. Clearly, our theory is not confining. 
Moreover, as shown Fig.~\ref{torelon} left, the combination $L m_T(L)$ is
approximately constant as $L$ is increased: the torelon mass varies
as $1/L$. Thus, there is no intrinsic mass scale which appears
in this channel: the torelon mass is set by the system size $L$.
This remarkable result is our first evidence that our theory is IR-conformal.
By relabelling the spatial direction $i$ as the imaginary time direction,
one realizes that we are measuring the correlation of two time-like Polyakov
loops, whose decay rate is governed by the Debye mass, given perturbatively 
at lowest order by $m_D(T) = 2 g T \sqrt{\frac{N_c}{3} + \frac{N_f}{6}}$.
This expression allows us to {\em define} a running coupling $g(L)$ via
\begin{align}
g(L) \equiv \frac{L m_T(L)}{2}\left(\frac{N_c}{3} + \frac{N_f}{6}\right)^{-1/2}
\end{align}
and we see that, in this scheme, our running coupling seems to go to a
non-zero constant as $L$ increases (although one cannot exclude, of course,
that it slowly goes to zero). Therefore, we have numerical evidence
supporting the view
that our strong-coupling, chirally symmetric theory is IR-conformal and
non-trivial.\footnote{
The extracted value of $g(L)$ approaches $g^*\sim 0.95$ and 
$\sim 0.80$ for $N_f=56$ and $96$ respectively. So $g^*$
decreases as $N_f$ increases: as $N_f$ 
keeps increasing, the ordering effect of the fermions increases, all
Wilson loops are driven towards 1. The theory 
becomes trivial for $N_f\to\infty$, even in the strong-coupling limit.
We come back to this point in Sec.~4.
}

$\bullet$ We now turn to the spectrum of the
Dirac operator.
We have analyzed the Dirac eigenvalue spectrum of the configurations in our
Monte Carlo ensembles, using a Lanczos algorithm to obtain an approximation
of the whole spectral density and an Arnoldi method to extract the smallest
eigenvalues to high accuracy.
The observable shown in Fig.~\ref{dirac} is the integrated eigenvalue density, 
defined as:
\begin{align}
\nu(\lambda) = \int_0^\lambda \rho(\bar{\lambda})d\bar{\lambda}=\frac{\rank(\lambda)}{\rank(\text{Dirac matrix})}  \in [0,1],
\end{align}
which counts the fraction of eigenvalues smaller than 
$\lambda$. Its derivative is simply the eigenvalue density $\rho(\lambda)$.
Comparison of this observable on quenched configurations ($\Nf=0$) 
with those in the chirally symmetric phase ($\Nf=56$), see Fig.~\ref{dirac} (left), yields qualitative differences of the spectra in the infrared:
The $\Nf=0$ curve starts linearly from the origin, 
reflecting an eigenvalue density approximately constant near $\lambda=0$.
On the contrary,
$\nu(\lambda)$ for $\Nf=56$ shows a spectral gap for small eigenvalues, implying $\rho(0)=0$,
which is consistent with chiral symmetry restoration according to the Banks-Casher relation.
The crucial question is on which scale does this spectral gap depend. 
To answer this question, in Fig.~\ref{dirac} (right) we compare $\nu(\lambda)$ at $\Nf=56$ for various lattice volumes, 
$L=4,6,8,10$ and $12$. We find that the spectral gap scales $\propto 1/L$ to a good approximation, 
which is a strong indication that our theory is IR-conformal\footnote{
If one would take the limits $L\to\infty$ first, then $m_q\to 0$, the expected
spectral density for a conformal theory would be $\rho(\lambda) \sim
\lambda^{(3-\gamma^*)/(1+\gamma^*)}$. Here, we take the opposite order of
limits.}: 
There does not seem to be any length scale in the chirally restored phase other than $L$.\footnote{
Small deviations from $1/L$ scaling allow us to 
extract the anomalous mass dimension $\gamma^*$. See \cite{JHEP}.
}

\section{The conjectured phase diagram}

\begin{figure}
\centerline{
\includegraphics[width=0.45\textwidth]{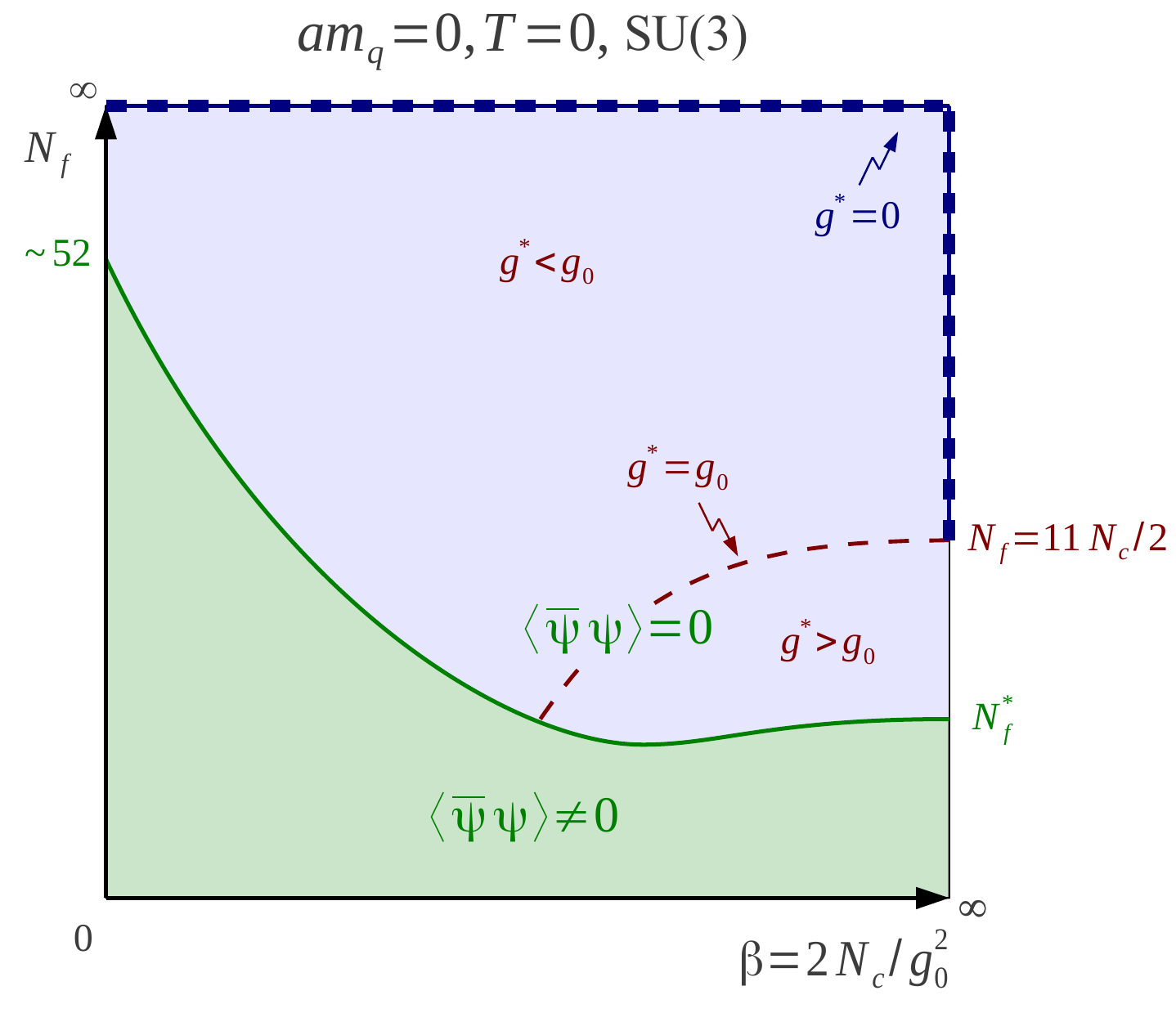}
\includegraphics[width=0.45\textwidth]{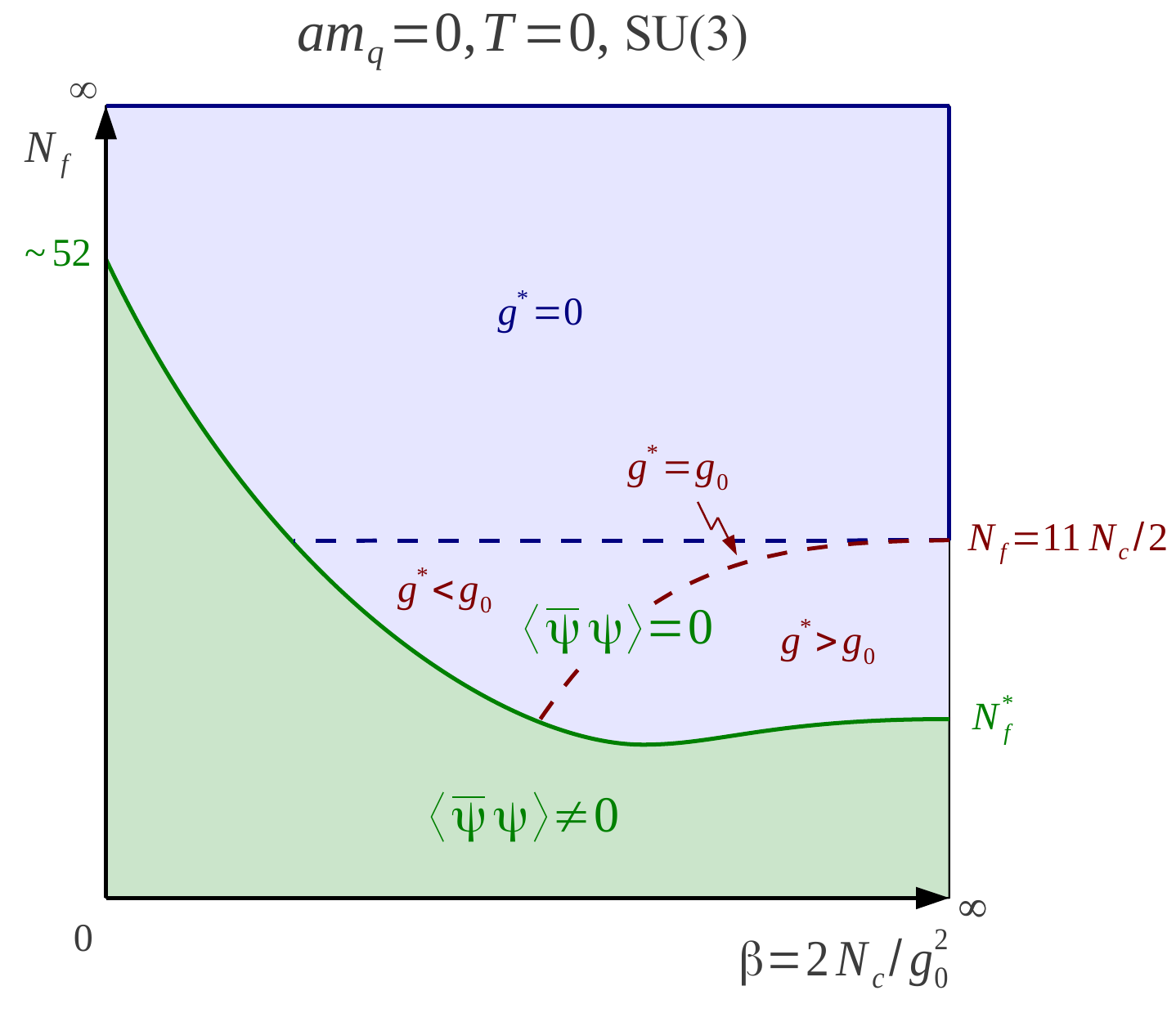}
}
\caption{Left: conjectured phase diagram in the $(\beta,\Nf)$ plane. A single
phase transition line separates the chirally broken phase from the chirally
symmetric, IR-conformal phase. The thick dotted line indicates trivial theories.
Right: alternative scenario, where the trivial theories extend to the area above $\Nf=11\Nc /2$. It is not favored by our measurements.}
\label{conjectured}
\end{figure}

We have presented evidence that the chirally restored phase at strong coupling 
is IR-conformal and non-trivial. In order to make the connection to the conformal window in the continuum, 
we now want to propose the phase diagram sketched in Fig.~\ref{conjectured} (left),
as a function of the plaquette coupling $\beta=6/g_0^2$ and of the number of
fundamental flavors $\Nf$.
Comparing our phase diagram with, e.g., those of 
Ref.~\cite{Banks-Zaks,Miransky-Yamawaki}, one can see substantial differences:
in our phase diagram, the $\beta=0$ IR-conformal 
phase is analytically connected with the weak-coupling, continuum IR-conformal 
phase - the simplest scenario supported by our exploratory scan
in $\beta$ as shown Fig.~\ref{fixedam} (right). A single first order phase transition line 
$\Nf^c(\beta)$ separates the region of broken chiral symmetry at small $\Nf$ 
from the chirally symmetric region at large $\Nf$.
Furthermore, based on our results for the running coupling Fig.~\ref{running_g},
we conjecture that for any finite $\beta$ and $\Nf>\Nf^c(\beta)$, 
large-$\Nf$ lattice QCD is IR-conformal, with a non-trivial fixed-point 
coupling $g^*>0$. This value changes continuously with $(\beta,\Nf)$, reaching 
the value zero for $\beta=\infty, \Nf > 33/2$ and for $\Nf=\infty~\forall 
\beta$,
as indicated Fig.~\ref{conjectured} (left) by a thick dotted line. $g^*$ grows
as one moves away from this dotted boundary towards the phase transition line.
An alternative scenario would be that the running coupling in Fig.~\ref{running_g}
slowly approaches zero, and $g^*=0$ for $N_f > 33/2$ for any
$\beta$ in the chirally symmetric phase. This is sketched in Fig.~\ref{conjectured} (right). 
In this scenario, the anomalous mass dimension at $\beta=0$
should vanish, $\gamma^*=0$, in contrast to our findings reported in \cite{JHEP}.
Finally, one may consider the dashed line $g^*=g_0$, with $g_0=(2\Nc/\beta)^{1/2}$, 
where the IR fixed-point coupling has the same value as the bare coupling.
This line starts at the point $(\beta=\infty,\Nf=33/2)$ where $g^*=g_0=0$.
Its precise location depends on the chosen renormalization scheme.
It is not associated with any kind of singularity of the free energy.

We have determined the phase diagram in the strong-coupling region only.
Studying the continuum limit is of course much more difficult, due to the 
large lattices that have to be used in order to control the finite size 
effects and the difficult control of lattice artifacts. 
We would like to suggest that the strong-coupling limit may represent
an advantageous ``poor man's laboratory'' for the study of $4d$ IR-conformal
gauge theories:
the range of scales over which conformal invariance applies
for a given computing effort is greatly enhanced at strong coupling.\footnote{At weak coupling, 
for a given lattice size $N^4$, the scales are ordered $a \ll 1/\Lambda \ll L=Na$,
where $\Lambda$ is the scale generated by the asymptotically free gauge 
dynamics. In contrast, at strong coupling, where the lattice becomes 
maximally coarse, the hierarchy is $a \sim 1/\Lambda \ll L=Na$.
Hence the dynamical range of conformal invariance, characterized by the product 
$L\Lambda$, is maximized at $\beta=0$ for a given lattice size $N=L/a$.}

\section{Conclusion}

We have shown that for $\beta=0$, contrary to common wisdom, there exists a 
strongly first-order, $\Nf$-driven bulk transition to a chirally symmetric phase. 
In the chiral limit, the transition occurs for $\Nf^c=52(4)$ \emph{continuum} flavors.
This finding is in contrast to the mean-field prediction, whose failure can
be traced back to approximations relying on $\Nf$ being small.
Clearly, the conventional, automatic association of the strong-coupling limit 
with confinement and chiral symmetry breaking is too naive.
Furthermore, the chirally restored phase extends to weak coupling.

We have also shown numerical evidence that the $\beta=0$ chirally restored phase of ``large-$\Nf$ QCD'' is IR-conformal, with a non-trivial,
$\Nf$-dependent value of the IR fixed-point coupling.
Since we have not observed any evidence for an additional $T=0$ phase transition as $\beta$ is increased, we speculate that the strong coupling chirally symmetric, IR-conformal phase is
analytically connected with the continuum IR-conformal phase.

One may ask how robust these statements are with respect to the particular
discretization of the Dirac operator and the gauge action. We think that the qualitative features will remain: chiral symmetry breaking
at strong coupling is a general consequence of the disorder in the gauge field. The ordering effect of fermions also is generic. So we
do expect an $\Nf$-driven bulk transition, generically of first-order. Actually, such a transition was observed
for Wilson fermions in Ref.~\cite{IwasakiNagai}. 
A first-order transition to a chirally broken phase as $\beta$ 
is reduced at fixed $\Nf$ has been observed many times for various lattice 
actions~\cite{Various}.
If the first-order nature persists all the way to the continuum limit,
then walking dynamics will not be observed, and the transition to the
conformal window will be characterized by ``jumping dynamics'', as proposed
by Sannino~\cite{Sannino2012}.

There are many directions in which to extend this exploratory study.
More observables, like the static potential, should be studied.\footnote{We have also studied the hadron spectrum: 
we find that parity partners are degenerate, and masses are exclusively due to finite-size effects: all hadron masses go to zero as the lattice size
$L$ is increased. Moreover, mass ratios such as $m_\pi/m_\rho$ should become constant.
Our results are reported in \cite{JHEP}.} 
One could study the $(\beta,\Nf)$ phase diagram as the gauge
group or the fermion content is changed. 
Also, and to make contact with other numerical
studies, a mass deformation can be introduced.\footnote{
We have studied the quark mass dependence of
the chiral condensate, but there are many technical difficutlies
associated with extracting the anomalous dimension $\gamma^*$, which we explain in \cite{JHEP}.}

\section{Acknowledgements}

We are grateful for stimulating discussions with P.~Damgaard, L.~Del Debbio, A.~Hasenfratz,\linebreak  
A.~Kovner, A.~Patella, K.~Rummukainen, F.~Sannino, K.~Splittorff, B.~Svetitsky and R.~Zwicky.
Computations have been carried out on the Brutus cluster at the ETH Z\"urich and on a small cluster in the Sejong University physics department.
S.~K.~is supported by Korea Foundation for International Cooperation of Science \& Technology (KICOS).
W.~U.~is supported by the Swiss National Science Foundation under grant 200020-122117.

\end{document}